# Photoioniziation of 3*d* electrons of Xe, Cs and Ba endohedral atoms: comparative analyses

**M. Ya. Amusia[1, 2], A. S. Baltenkov[3] and L. V. Chernysheva[2]**

[1]Racah Institute of Physics, The Hebrew University, Jerusalem 91904, Israel
[2]Ioffe Physical-Technical Institute, St.-Petersburg 194021, Russia
[3]Arifov Institute of Electronics, Tashkent, 700125, Uzbekistan

**Abstract**

We demonstrate rather interesting manifestations of co-existence of resonance features in characteristics of the photoionization of 3*d*-electrons in Xe, Cs and Ba endohedral atoms. It is shown that for all of the considered atoms the reflection by the fullerene shell of photoelectrons produced by the 3*d* subshell photoionization affects greatly partial photoionization cross-sections of $3d_{5/2}$ and $3d_{3/2}$ levels and respective angular anisotropy parameters, both dipole and non-dipole adding to all of them additional maximums and minimums.

The results obtained demonstrate distinctive differences between the three atoms. The calculations are performed treating the 3/2 and 5/2 electrons as electrons of different kinds with their spins "up" and "down". The effect of $C_{60}$ shell is accounted for in the frame of the "orange" skin potential model. It is essential that in the considered photon frequency region presented resonance features are not affected by the $C_{60}$ polarization.

## 1. Introduction

Recently we have published the results of calculations of the photoionization cross-section for 3*d* electrons in Xe@$C_{60}$ where along with intra-doublet resonance there is exist a strong action of the fullerene $C_{60}$ potential upon the electron waves that are emitted from both, $3d_{3/2}$ and $3d_{5/2}$ levels [1]. This leads to interference patterns of the photoionization cross-section and angular anisotropy parameters, both dipole and non-dipole. The intra-doublet resonance in 3*d* is located at high enough photon energies, where the role of fullerene polarization by the incoming radiation [2] is entirely inessential.

We know since the studies of intra-doublet resonances in isolated atoms that it is an essential difference in their manifestations in three atoms-neighbors – Xe, Cs, and Ba. While in Xe this resonance is an entirely continuous spectrum phenomenon, it acquires an autoionization nature in Ba. As a result, all characteristics of the photoionization process, namely the partial cross-sections and angular anisotropy parameters are essentially different for 3*d* in this sequence of elements [3].

Similarly, the photoionization of 3*d* -electrons in Cs@$C_{60}$ and Ba@$C_{60}$ could be essentially different from that of Xe@$C_{60}$. To investigate this difference is the aim of the present paper. We will present newly obtained data for Cs@$C_{60}$ and Ba@$C_{60}$ and compare it with that for Xe@$C_{60}$.

To justify this investigation and to put it into the context of the current research in the field under consideration, let us repeat some information that was presented in the introduction to [1]. This will simplify considerably the understanding of the aims and expected results of the current research.

A great deal of attention during last years was (and still is) concentrated on photoionization of endohedral atoms. It was demonstrated in a number of papers [4-12] that the fullerenes shell adds prominent resonance structure in the photoionization cross section of endohedral atoms. Although the experimental investigation of A@$C_{60}$ photoionization seems to be very difficult at this moment, it will be inevitably studied in the future. First measurements for such objects are already performed [13]. This justifies and stimulates the current efforts of the theorists that are predicting rather non-trivial effects waiting for verification.

The role of $C_{60}$ in A@$C_{60}$ photoionization is manifold. $C_{60}$ act as a spherical potential resonator that reflects the photoelectron wave coming from A atom. This leads to interference of out-coming and in-coming (reflected) waves and to confinement resonances in the frequency dependence of the photoionization cross sections [8]. These resonances were studied in [4-7] for the example of endohedral Ba and Ca atoms located inside the $C_{60}$ shell. Valence and core photoemission of endohedral atoms A@$C_{60}$ (A=Li, Na, K, Be, Mg, and Ca) and resonances in the total cross section were also studied in [8, 11] .

The interference of the photoelectron spherical waves inside the resonator $C_{60}$ affects significantly not only the total cross section but also the angular distribution of photoelectrons. This phenomenon was analysed in [9], where it was shown that the confinement resonances are found also in the frequency dependencies of the dipole and nondipole parameters of the photoelectron angular distribution. The results of these studies give evidence that the reflection and refraction of the photoelectron waves by the potential resonator $C_{60}$ is prominent up to 60 – 80 eV of the electron energy.

The $C_{60}$ shell at some frequencies acts as a dynamical screen that is capable to suppress or enhance the incident electromagnetic radiation acting upon the doped atom A [14-16]. This effect is due to dynamical polarization of the collectivized electrons of the fullerene shell. Plasma excitations of these electrons generate the alternating dipole moment. This dipole moment causes the ionization of the electronic shells of the endohedral atom. The screening effects of the $C_{60}$ shell are particularly strong for incident radiation frequency $\omega$ */ of about that of the $C_{60}$ Giant resonance, i.e. 20 – 22 eV, but is noticeable in a much broader region, from ionization threshold up to 60 – 80 eV.

The resonator and screen effects of the fullerene shell $C_{60}$ manifest themselves as resonance structure in the frequency dependencies of the differential and total cross sections of the endohedral atom photoionization. Of interest are the situations when pure atomic resonances are superimposed with resonances connected to reflection and refraction of atomic photoelectrons by $C_{60}$ shell and modification of the incoming photon by the resonating fulleren shell.

We have studied before [15, 17] the effect of $C_{60}$ upon the most important atomic resonances, the Giant and interference, both in Xe. However, these resonances are located at relatively low photon energies, at about 80-100 eV for the Giant and 30-100 for the interference resonance. This is why they are affected by both the reflection from $C_{60}$ shell and by the fullerene own Giant resonance that is located at about 20 eV.

However, a prominent atomic resonance exists at sufficiently high energies. It was called intra-doublet resonance [2, 18]. It was observed in photoionization of $3d_{5/2}$ and $3d_{3/2}$ electrons in Xe [3] and in Cs [19] and interpreted and discussed in [2, 18].

---

* / Atomic system of units is used in this paper

Les us remind that the intra-doublet resonance is a result of action of $3d_{3/2}$ electrons upon $3d_{5/2}$. The threshold energy of $3d$ is almost 700 eV in Xe and even a little bit higher in Cs and Ba, i.e. well above the position of the Giant resonance in $C_{60}$. The interference between the intra-doublet and confinement resonances in Xe@$C_{60}$ was discussed in [1]. Here we will see similar interference in Cs@$C_{60}$ and Ba@$C_{60}$. As in the Xe case, at these energies the action of confinement resonances upon atomic became clear and not disturbed by the virtual or real excitations of $C_{60}$.

## 2. Main formulas

The theoretical approach employed in this paper is similar to that used in [1]. For convenience of our readers, we will repeat it s main points again. Let us start with the problem of an isolated atom. In fact, what we present below is not a derivation but an extended explanation of notation used.

The method to treat the result of mutual action of $3d_{5/2}$ and $3d_{3/2}$ electrons for isolated atom Xe was discussed as two semi-filled levels with five spin-*up* $(\uparrow)$ and five spin-*down* $(\downarrow)$ electrons each was presented for the first time in [18]. Then the Random Phase Approximation with Exchange (RPAE) equations for atoms with semi-filled shells (so-called Spin Polarized RPAE or SP RPAE) are solved, as described in e.g. [20]. For semi-filled shell atoms the following relation gives the differential in angle photoionization cross-section by non-polarized light, which is similar to that of closed shell atoms [21] (see also e.g. [22]):

$$\frac{d\sigma_{nl\uparrow\downarrow}(\omega)}{d\Omega} = \frac{\sigma_{nl\uparrow\downarrow}(\omega)}{4\pi}[1 - \frac{\beta_{nl\uparrow\downarrow}}{2} P_2(\cos\theta) + \kappa\gamma_{nl\uparrow\downarrow} P_1(\cos\theta) + \kappa\eta_{nl\uparrow\downarrow} P_3(\cos\theta)], \qquad (1)$$

where $\kappa = \omega / c$, $P_l(\cos\theta)$ are the Legendre polynomials, $\theta$ is the angle between photon $\boldsymbol{\kappa}$ and photoelectron velocity $\mathbf{v}$, $\beta_{nl\uparrow\downarrow}(\omega)$ is the dipole, while $\gamma_{nl\uparrow\downarrow}(\omega)$ and $\eta_{nl\uparrow\downarrow}(\omega)$ are so-called non-dipole angular anisotropy parameters, where the arrows $\uparrow\downarrow$ relates the corresponding parameters to *up* and *down* electrons, respectively.

Since in experiment, usually the sources of linearly polarized radiation are used, instead of (1) another form of angular distribution is more convenient [23, 24]:

$$\frac{d\sigma_{nl\uparrow\downarrow}(\omega)}{d\Omega} = \frac{\sigma_{nl\uparrow\downarrow}(\omega)}{4\pi}[1 + \beta_{nl\uparrow\downarrow} P_2(\cos\Theta) + (\delta^C_{nl\uparrow\downarrow} + \gamma^C_{nl\uparrow\downarrow}\cos^2\Theta)\sin\Theta\cos\Phi]. \qquad (2)$$

Here $\Theta$ is the polar angle between the vectors of photoelectron's velocity $\mathbf{v}$ and photon's polarization $\mathbf{e}$, while $\Phi$ is the azimuth angle determined by the projection of $\mathbf{v}$ in the plane orthogonal to $\mathbf{e}$ that includes the vector of photon's velocity. The non-dipole parameters in (1) and (2) are connected by the simple relations [22]

$$\frac{\gamma^C_{nl\uparrow\downarrow}}{5} + \delta^C_{nl\uparrow\downarrow} = \kappa\gamma_{nl\uparrow\downarrow}, \qquad \frac{\gamma^C_{nl\uparrow\downarrow}}{5} = -\kappa\eta_{nl\uparrow\downarrow}. \qquad (3)$$

The below-presented results of calculations of non-dipole parameters are obtained using both expressions (1) and (2). There are two possible dipole transitions from subshell *l*, namely $l \to l \pm 1$ and three quadrupole transitions $l \to l; l \pm 2$. Corresponding general expressions for $\beta_{nl\uparrow\downarrow}(\omega)$, $\gamma_{nl\uparrow\downarrow}(\omega)$ and $\eta_{nl\uparrow\downarrow}(\omega)$ are rather complex and content the dipole $d_{l\pm1}$ and quadrupole $q_{l\pm2,0}$ matrix elements of photoelectron transitions. In one-electron Hartree-Fock (HF) approximation these parameters can be presented as [22, 25]:

$$\beta_{nl\uparrow\downarrow}(\omega)=\frac{1}{(2l+1)\left[(l+1)d^2_{l+1\uparrow\downarrow}+ld^2_{l-1\uparrow\downarrow}\right]}[(l+1)(l+2)d^2_{l+1\uparrow\downarrow}+l(l-1)d^2_{l-1\uparrow\downarrow}-$$
$$6l(l+1)d_{l+1\uparrow\downarrow}d_{l-1\uparrow\downarrow}\cos(\delta_{l+1}-\delta_{l-1})]\,. \qquad (4)$$

It is implied that the indexes ↑↓ are added similarly to the parameters $\gamma_{nl}(\omega)$, $\eta_{nl}(\omega)$ and matrix elements $d_{l\pm1}$, $q_{l,l\pm2}$ in (5) and (6):

$$\gamma_{nl}(\omega)=\frac{3}{5\left[ld^2_{l-1}+(l+1)d^2_{l+1}\right]}\left\{\frac{l+1}{2l+3}[3(l+2)q_{l+2}d_{l+1}\cos(\delta_{l+2}-\delta_{l+1})-lq_l d_{l+1}\times\right.$$
$$\left.\times\cos(\delta_{l+2}-\delta_{l+1})]-\frac{l}{2l+1}[3(l-1)q_{l-2}d_{l-1}\cos(\delta_{l-2}-\delta_{l-1})-(l+1)q_l d_{l-1}\cos(\delta_l-\delta_{l-1})]\right\}, \qquad (5)$$

$$\eta_{nl}(\omega)=\frac{3}{5\left[ld^2_{l-1}+(l+1)d^2_{l+1}\right]}\left\{\frac{(l+1)(l+2)}{(2l+1)(2l+3)}q_{l+2}\left[5ld_{l-1}\cos(\delta_{l+2}-\delta_{l-1})-\right.\right.$$
$$\left.-(l+3)d_{l+1}\cos(\delta_{l+2}-\delta_{l-1})\right]-\frac{(l-1)l}{(2l+1)(2l+1)}q_{l-2}\times$$
$$\times\left[5(l+1)d_{l+1}\cos(\delta_{l-2}-\delta_{l+1})-(l-2)d_{l-1}\cos(\delta_{l-2}-\delta_{l-1})\right]+$$
$$\left.+2\frac{l(l+1)}{(2l-1)(2l+3)}q_l\left[(l+2)d_{l+1}\cos(\delta_l-\delta_{l+1})-(l-1)d_{l-1}\cos(\delta_l-\delta_{l-1})\right]\right\}. \qquad (6)$$

Here $\delta_l(k)$ are the photoelectrons' scattering phases; the following relation gives the matrix elements $d_{l\pm1\uparrow\downarrow}$ in the so-called *r*-form

$$d_{l\pm1\uparrow\downarrow}\equiv\int_0^\infty P_{nl\uparrow\downarrow}(r)rP_{\varepsilon l\pm1\uparrow\downarrow}(r)dr, \qquad (7)$$

where $P_{nl\uparrow\downarrow}(r)$, $P_{\varepsilon l\pm1\uparrow\downarrow}(r)$ are the radial so-called Spin-Polarized Hartree-Fock (SP HF) [26] one-electron wave functions of the *nl* discrete level and $\varepsilon\, l\pm1$ - in continuous spectrum, respectively. The following relation gives the quadrupole matrix elements

$$q_{l\pm2,0\updownarrow}\equiv\frac{1}{2}\int_0^\infty P_{nl\updownarrow}(r)r^2P_{\varepsilon l\pm2,0\updownarrow}(r)dr. \qquad (8)$$

In order to take into account the Random Phase Approximation with Exchange (RPAE) [25] multi-electron correlations, one has to perform the following substitutions in the expressions for $\beta_{nl\uparrow\downarrow}(\omega)$, $\gamma_{nl\uparrow\downarrow}(\omega)$ and $\eta_{nl\uparrow\downarrow}(\omega)$ [22]:

$$d_{l+1}d_{l-1}\cos(\delta_{l+1}-\delta_{l-1})\to\left[\left(\operatorname{Re}D_{l+1}\operatorname{Re}D_{l-1}+\operatorname{Im}D_{l+1}\operatorname{Im}Q_{l\pm2,0}\right)\cos\left(\delta_{l\pm2,0}-\delta_{l\pm1}\right)-\right.$$
$$\left.-\left(\operatorname{Re}D_{l+1}\operatorname{Im}D_{l-1}-\operatorname{Im}D_{l+1}\operatorname{Re}Q_{l-1}\right)\sin\left(\delta_{l+1}-\delta_{l-1}\right)\right]\equiv \tag{9}$$
$$\tilde{D}_{l+1}\tilde{D}_{l-1}\cos\left(\delta_{l\pm2,0}+\Delta_{l\pm2,0}-\delta_{l\pm1}-\Delta_{l\pm1}\right).$$

$$d_{l\pm1}q_{l\pm2.0}\cos\left(\delta_{l\pm2,0}-\delta_{l\pm1}\right)\to\left[\left(\operatorname{Re}D_{l\pm1}\operatorname{Re}Q_{l\pm2,0}+\operatorname{Im}D_{l\pm1}\operatorname{Im}Q_{l\pm2,0}\right)\cos\left(\delta_{l\pm2,0}-\delta_{l\pm1}\right)-\right.$$
$$\left.-\left(\operatorname{Re}D_{l\pm1}\operatorname{Im}Q_{l\pm2,0}-\operatorname{Im}D_{l\pm1}\operatorname{Re}Q_{l\pm2,0}\right)\sin\left(\delta_{l\pm2,0}-\delta_{l\pm1}\right)\right]\equiv$$
$$\tilde{D}_{l\pm1}\tilde{Q}_{l\pm2,0}\cos\left(\delta_{l\pm2,0}+\Delta_{l\pm2,0}-\delta_{l\pm1}-\Delta_{l\pm1}\right), \tag{10}$$
$$d_{l\pm1}^2\to\operatorname{Re}D_{l\pm1}^2+\operatorname{Im}D_{l\pm1}^2\equiv\tilde{D}_{l\pm1}^2.$$

Here the following notations are used for the matrix elements with account of multi-electron correlations, dipole and quadrupole, respectively:

$$D_{l\pm1}(\omega)\equiv\tilde{D}_{l\pm1}(\omega)\exp[i\Delta_{l\pm1}(\varepsilon)];\ Q_{l\pm2,0}(\omega)\equiv\tilde{Q}_{l\pm2,0}(\omega)\exp[i\Delta_{l\pm2,0}(\varepsilon)], \tag{11}$$

where $\tilde{D}_{l\pm1}(\omega)$, $\tilde{Q}_{l\pm2,0}(\omega)$, $\Delta_{l\pm1}$ and $\Delta_{l\pm2,0}$ are absolute values of the amplitudes for respective transitions and phases for photoelectrons with angular moments $l\pm1$ and $l\pm2,0$.

The following are the ordinary RPAE equation for the dipole matrix elements

$$\langle\nu_2|D(\omega)|\nu_1\rangle=\langle\nu_2|d|\nu_1\rangle+\sum_{\nu_3,\nu_4}\frac{\langle\nu_3|D(\omega)|\nu_4\rangle\left(n_{\nu_4}-n_{\nu_3}\right)\langle\nu_4\nu_2|U|\nu_3\nu_1\rangle}{\varepsilon_{\nu_4}-\varepsilon_{\nu_3}+\omega+i\eta\left(1-2n_{\nu_3}\right)}, \tag{12}$$

where

$$\langle\nu_1\nu_2|\hat{U}|\nu_1'\nu_2'\rangle\equiv\langle\nu_1\nu_2|\hat{V}|\nu_1'\nu_2'\rangle-\langle\nu_1\nu_2|\hat{V}|\nu_2'\nu_1'\rangle. \tag{13}$$

Here $\hat{V}\equiv 1/|\vec{r}-\vec{r}'|$ and $\nu_i$ is the total set of quantum numbers that characterize a HF one-electron state on discrete (continuum) levels. That includes the principal quantum number (energy), angular momentum, its projection and the projection of the electron spin. The function $n_{\nu_i}$ (the so-called step-function) is equal to 1 for occupied and 0 for vacant states.

For semi-filled shells the RPAE equations are transformed into the following system of equations that can be presented in the matrix form:

$$\left(\hat{D}_\uparrow(\omega)\hat{D}_\downarrow(\omega)\right)=\left(\hat{d}_\uparrow(\omega)\hat{d}_\downarrow(\omega)\right)+\left(\hat{D}_\uparrow(\omega)\hat{D}_\downarrow(\omega)\right)\times\begin{pmatrix}\hat{\chi}_{\uparrow\uparrow} & 0\\ 0 & \hat{\chi}_{\downarrow\downarrow}\end{pmatrix}\times\begin{pmatrix}\hat{U}_{\uparrow\uparrow} & \hat{V}_{\uparrow\downarrow}\\ \hat{V}_{\downarrow\uparrow} & \hat{U}_{\downarrow\downarrow}\end{pmatrix}. \tag{14}$$

The dipole matrix elements $D_{l\pm1}$ are obtained by solving the radial part of the RPAE equation (12). As to the quadrupole matrix elements $Q_{l\pm2,0}$, they are obtained by solving the radial part of the RPAE equation, similar to (12)

$$\langle \nu_2 | Q(\omega) | \nu_1 \rangle = \langle \nu_2 | \hat{q} | \nu_1 \rangle + \sum_{\nu_3,\nu_4} \frac{\langle \nu_3 | Q(\omega) | \nu_4 \rangle (n_{\nu_4} - n_{\nu_3}) \langle \nu_4 \nu_2 | U | \nu_3 \nu_1 \rangle}{\varepsilon_{\nu_4} - \varepsilon_{\nu_3} + \omega + i\eta (1 - 2n_{\nu_3})}. \quad (15)$$

Here in *r*-form one has $\hat{q} = r^2 P_2(\cos\theta)$.

Equations (14, 15) are solved numerically using the procedure discussed at length in [25]. The generalization of (15) for semi-filled shells is similar to (14):

$$\left(\hat{Q}_{\uparrow}(\omega)\hat{Q}_{\downarrow}(\omega)\right) = \left(\hat{q}_{\uparrow}(\omega)\hat{q}_{\downarrow}(\omega)\right) + \left(\hat{Q}_{\uparrow}(\omega)\hat{Q}_{\downarrow}(\omega)\right) \times \begin{pmatrix} \hat{\chi}_{\uparrow\uparrow} & 0 \\ 0 & \hat{\chi}_{\downarrow\downarrow} \end{pmatrix} \times \begin{pmatrix} \hat{U}_{\uparrow\uparrow} & \hat{V}_{\uparrow\downarrow} \\ \hat{V}_{\downarrow\uparrow} & \hat{U}_{\downarrow\downarrow} \end{pmatrix} \quad (16)$$

where

$$\hat{\chi}(\omega) = \hat{1}/(\omega - \hat{H}_{ev}) - \hat{1}/(\omega + \hat{H}_{ev}). \quad (17)$$

In (17) $\hat{H}_{ev}$ is the electron – vacancy HF Hamiltonian. Equations (14) and (16) permit to treat the electron correlations of $3d_{5/2}$ and $3d_{3/2}$ levels, if to corrected (14) and (16) by adding factors 6/5 and 4/5 to the *up* and *down* states, respectively. These equations were solved in this paper numerically. The cross-sections and angular anisotropy parameters are calculated by using numerical procedures with the codes described in [26].

### 3. Effect of $C_{60}$ fullerene shell

The ionization potentials of the Xe $3d_{5/2}$ and $3d_{3/2}$ subshells have the value of about 50 Ry. For this photon energy significantly exceeding the energy of the Giant Resonance of the $C_{60}$ collectivised electrons, the screening effects of the fullerene shell do not play any essential role in the photoionization processes of the Xe atom localized inside $C_{60}$. Therefore, the main reason for modifying the photoionization parameters of $3d_{5/2}$ and $3d_{3/2}$ subshells of the encapsulated Xe, as compared to the free Xe, are the confinement effects. These effects near the photoionization threshold can be described within the framework of the "orange" skin potential model. According to this model, for small photoelectron energies the real static potential of the $C_{60}$ can be presented by a zero-thickness bubble pseudo-potential (see [27, 28] and references therein):

$$V(r) = -V_0 \delta(r - R). \quad (18)$$

The parameter $V_0$ is determined by the requirement that the binding energy of the extra electron in the negative ion $C_{60}^-$ is equal to its observable value. Addition of potential (18) to the atomic HF potential leads to a factor $F_l(k)$ in the photoionization amplitudes which depends only upon the photoelectron's momentum $k$ and orbital quantum number $l$ [28]:

$$F_l(k) = \cos\Delta_l(k) \left[ 1 - \tan\Delta_l(k) \frac{v_{kl}(R)}{u_{kl}(R)} \right], \quad (19)$$

where $\Delta_l(k)$ are the additional phase shifts due to the fullerene shell potential (18). They are expressed by the following formula:

$$\tan\Delta_l(k) = \frac{u_{kl}^2(R)}{u_{kl}(R)v_{kl}(R) + k/2V_0}. \qquad (20)$$

In these formulas $u_{kl}(r)$ and $v_{kl}(r)$ are the regular and irregular solutions of the atomic HF equations for a photoelectron with momentum $k = \sqrt{2\varepsilon}$, where $\varepsilon$ is the photoelectron energy connected with the photon energy $\omega$ by the relation $\varepsilon = \omega - I_A$ with $I$ being the atom A ionization potential.

Using Eq. (19), one can obtain the following relation for $D^{AC}$ and $Q^{AC}$ amplitudes of endohedral atom expressed via the respective values for isolated atom that correspond to $nl \to \varepsilon l'$ transitions:

$$D(Q)_{nl,kl'}^{AC}(\omega) = F_{l'}(\omega) D(Q)_{nl,kl'}(\omega). \qquad (21)$$

Factors $F_{l'}(\omega)$ are complex numbers

$$F_{l'}(\omega) = \tilde{F}_{l'}(\omega)\exp[i\Lambda_{l'}(\varepsilon)] \equiv \tilde{F}_{l'}\exp[i\Lambda_{l'}]. \qquad (22)$$

For the cross-sections one has

$$\sigma_{nl,kl'}^{AC}(\omega) = |F_{l'}(\omega)|^2 \sigma_{nl,kl'}^{A}(\omega) \equiv \tilde{F}_{l'}(\omega)^2 \sigma_{nl,kl'}^{A}(\omega). \qquad (23)$$

With these amplitudes, using the expressions (4-6) and performing the substitution (9, 10) we obtain the cross-sections for Xe@$C_{60}$ and angular anisotropy parameters. While calculating the anisotropy parameters, the cosines of atomic phases differences $\cos(\delta_l - \delta_{l'})$ in formulas (4)-(6) are replaced by $\cos(\delta_l + \Delta_l - \delta_{l'} - \Delta_{l'})$. Note that factors $F_{l'}(\omega)$ are different for spin-*up* $(\uparrow)$ and spin-*down* $(\downarrow)$ photoelectrons and therefore have to be denoted as $F_{l'\uparrow\downarrow}(\omega)$.

As a result, one has for the dipole angular anisotropy parameter (4), using (9) and (10):

$$\beta_{nl\uparrow\downarrow}(\omega) = \frac{1}{(2l+1)\left[(l+1)\tilde{F}_{l+1\uparrow\downarrow}^2 \tilde{D}_{l+1\uparrow\downarrow}^2 + l\tilde{F}_{l-1\uparrow\downarrow}^2 \tilde{D}_{l-1\uparrow\downarrow}^2\right]}[(l+1)(l+2)\tilde{F}_{l+1\uparrow\downarrow}^2 \tilde{D}_{l+1\uparrow\downarrow}^2$$
$$+ l(l-1)\tilde{F}_{l-1\uparrow\downarrow}^2 \tilde{D}_{l-1\uparrow\downarrow}^2 - 6l(l+1)\tilde{F}_{l+1\uparrow\downarrow}\tilde{F}_{l-1\uparrow\downarrow}\tilde{D}_{l+1\uparrow\downarrow}\tilde{D}_{l-1\uparrow\downarrow}\cos(\tilde{\delta}_{l+1} - \tilde{\delta}_{l-1})] \qquad . \qquad (24)$$

where $\tilde{\delta}_{l'} = \delta_{l'} + \Delta_{l'} + \Lambda_{l'}$ (see (11) and (22)).

From (5) and (6), using (9) and (10) we arrive to the following expressions for the non-dipole angular anisotropy parameters:

$$\gamma_{nl}(\omega)=\frac{3}{5\left[(l+1)\tilde{F}^2_{l+1\uparrow\downarrow}\tilde{D}^2_{l+1\uparrow\downarrow}+l\tilde{F}^2_{l-1\uparrow\downarrow}\tilde{D}^2_{l-1\uparrow\downarrow}\right]}\times$$

$$\times\left\{\frac{(l+1)\tilde{F}_{l+1\uparrow\downarrow}}{2l+3}[3(l+2)\tilde{F}_{l+2\uparrow\downarrow}\tilde{Q}_{l+2\uparrow\downarrow}\tilde{D}_{l+1\uparrow\downarrow}\cos(\tilde{\delta}_{l+2}-\tilde{\delta}_{l+1})-l\tilde{F}_{l\uparrow\downarrow}\tilde{Q}_{l\uparrow\downarrow}\tilde{D}_{l+1\uparrow\downarrow}\cos(\tilde{\delta}_{l+2}-\tilde{\delta}_{l+1})]-\right.$$

$$\left.-\frac{l\tilde{F}_{l-1\uparrow\downarrow}}{2l+1}\left[3(l-1)\tilde{F}_{l-2\uparrow\downarrow}\tilde{Q}_{l-2\uparrow\downarrow}\tilde{D}_{l-1\uparrow\downarrow}\cos(\tilde{\delta}_{l-2}-\tilde{\delta}_{l-1})-(l+1)\tilde{F}_{l\uparrow\downarrow}\tilde{Q}_{l\uparrow\downarrow}\tilde{D}_{l-1\uparrow\downarrow}\cos(\tilde{\delta}_{l}-\tilde{\delta}_{l-1})\right]\right\}$$

(25)

$$\eta_{nl}(\omega)=\frac{3}{5\left[(l+1)\tilde{F}^2_{l+1\uparrow\downarrow}\tilde{D}^2_{l+1\uparrow\downarrow}+l\tilde{F}^2_{l-1\uparrow\downarrow}\tilde{D}^2_{l-1\uparrow\downarrow}\right]}\times$$

$$\left\{\frac{(l+1)(l+2)}{(2l+1)(2l+3)}\tilde{F}_{l+2\uparrow\downarrow}\tilde{Q}_{l+2\uparrow\downarrow}\left[5l\tilde{F}_{l-1\uparrow\downarrow}\tilde{D}_{l-1\uparrow\downarrow}d_{l-1}\cos(\tilde{\delta}_{l+2}-\tilde{\delta}_{l-1})-\right.\right.$$

$$\left.-(l+3)\tilde{F}_{l+1\uparrow\downarrow}\tilde{D}_{l+1\uparrow\downarrow}\cos(\tilde{\delta}_{l+2}-\tilde{\delta}_{l-1})\right]-\frac{(l-1)l}{(2l+1)(2l+1)}\tilde{F}_{l-2\uparrow\downarrow}\tilde{Q}_{l-2\uparrow\downarrow}\times$$

$$\times\left[5(l+1)\tilde{F}_{l+1\uparrow\downarrow}\tilde{D}_{l+1\uparrow\downarrow}\cos(\tilde{\delta}_{l-2}-\tilde{\delta}_{l+1})-(l-2)\tilde{F}_{l-1\uparrow\downarrow}\tilde{D}_{l-1\uparrow\downarrow}\cos(\tilde{\delta}_{l-2}-\tilde{\delta}_{l-1})\right]+$$

$$\left.+2\frac{l(l+1)\tilde{F}_{l\uparrow\downarrow}\tilde{Q}_{l\uparrow\downarrow}}{(2l-1)(2l+3)}\left[(l+2)\tilde{F}_{l+1\uparrow\downarrow}\tilde{D}_{l+1\uparrow\downarrow}\tilde{D}_{l+1}\cos(\tilde{\delta}_{l}-\tilde{\delta}_{l+1})-(l-1)\tilde{F}_{l-1\uparrow\downarrow}\tilde{D}_{l-1\uparrow\downarrow}\tilde{D}_{l-1}\cos(\tilde{\delta}_{l}-\tilde{\delta}_{l-1})\right]\right\}.$$

(26)

## 4. Some details of calculations and their results

Naturally, the $C_{60}$ parameters in the present calculations were chosen the same as in the previous papers, e.g. in [27, 28]: $R = 6.639$ and $V_0 = 0.443$. In Fig. 1-5 we present our results for the partial cross-sections $\sigma^A_{nl,kl'}(\omega)$ that correspond to $3d \rightarrow \varepsilon f$ and $3d \rightarrow \varepsilon p$ transitions, dipole and non-dipole angular anisotropy parameters for Xe, Cs and Ba endohedral atoms. The results for Xe@$C_{60}$ were already presented in [1], while the data for isolated Xe can be found in [18][14]. The photoionization cross-section and angular anisotropy parameters for $3d$ electrons in Cs and Ba were considered in [29].

The partial cross-sections for $3d \rightarrow \varepsilon f$ and $3d \rightarrow \varepsilon p$ transitions in Xe, Cs and Ba are presented in Fig. 1a, b, c, in its upper and lower parts, respectively. The solid lines in these figures present the $3d_{5/2}$ cross sections while the dashed line stands for the $3d_{3/2}$ cross-sections. The contribution of the transition $3d \rightarrow \varepsilon f$ exceeds that of $3d \rightarrow \varepsilon p$ by almost two orders of magnitude. An additional maximum in both transitions from the $3d_{5/2}$ level is due to action of the electrons on the $3d_{3/2}$ level. Since the ionization potentials of the "up" and "down" levels are different the corresponding curves are shifted relative to each other along the axis of photon energy $\omega$. The main maxima in the $3d \rightarrow \varepsilon f$ transition are increasing on the way from Xe to Ba. In Ba it is a result of autoionization of the discrete excitation of $3d_{3/2}$ electrons.

Since the partial cross-section corresponding to the $3d \rightarrow \varepsilon p$ transition is much smaller than the $3d \rightarrow \varepsilon f$ partial cross-section, when considering the effect of the fullerene shell on the Xe@$C_{60}$ photoionization cross-section we concentrate only on the main $3d \rightarrow \varepsilon f$ electron transition.

In Fig. 2a, b, c we present the partial photoionization cross-section of $3d_{5/2}$ and $3d_{3/2}$ levels in Xe@$C_{60}$, Cs@$C_{60}$, and Ba@$C_{60}$. The solid lines in these figures depict the results for endohedral atoms, while the dashed line presents the same for isolated atoms. The upper part of Fig. 2a, b, c demonstrates the effect of $C_{60}$ upon the photoionization cross-section of the $3d_{5/2}$ electrons, while the lower parts present results for $3d_{3/2}$ electrons. The lower part is the cross-section of the $3d_{3/2}$ electrons. The main maximum in Xe@$C_{60}$ and Cs@$C_{60}$ remains almost the same, while in Ba@$C_{60}$ it is considerably smaller. All cross-sections acquire a prominent oscillating structure around the background curves, due to reflection of the $\varepsilon f$ photoelectron wave by the $C_{60}$ shell.

Note that the additional maximum that appears due to the $3d_{3/2}$ action is prominently altered under the influence of the $C_{60}$ shell. Several additional secondary maxima are created. The complex oscillating structure makes the cross-sections of 5/2 and 3/2 electrons in Xe@$C_{60}$ similar. To a lesser extent this similarity takes place also in Cs@$C_{60}$. In Ba@$C_{60}$ the amplitude of oscillations is even bigger for the 3/2 cross-section.

The amplitude factor Eq. (18) is defined by the values of the photoelectron wave functions with the "up" and "down" spins at a point $r = R$. Since we deal with the completely filled subshell the role of the spin effects in the wave function behaviour is small. Hence the continuum wave functions for the "up" and "down" spins at this point are almost equal. So, the amplitude factors $F_l(k)$ for the 3/2 and 5/2 levels are similar to each other. However, since the background cross-sections for the isolated Xe, Cs and Ba atoms are significantly different, the interference of 3/2 and 5/2 intra-doublet resonances and confinement resonances leads to essentially different cross-sections for Xe@$C_{60}$, Cs@$C_{60}$, and Ba@$C_{60}$ compounds, which we see in Fig. 2a, b, c.

In Fig. 3a, b, c we depict the dipole angular anisotropy parameters $\beta(\omega)$ for the endohedrals Xe@$C_{60}$, Cs@$C_{60}$, and Ba@$C_{60}$ in the comparison of the isolated atoms Xe, Cs and Ba, respectively. Impressive modifications are seen. The solid lines in these figures depict the results for endohedral atoms, while the dashed line presents the same for isolated atoms. The upper part of Fig. 3a, b, c demonstrates the effect of $C_{60}$ upon the photoionization cross-section of the $3d_{5/2}$ electrons. The lower part presents the results for $3d_{3/2}$ electrons. Although being an order of magnitude smaller than the $3d \to \varepsilon d$ amplitude, the $3d \to \varepsilon p$ amplitude is much more important in the angular distributions of photoelectrons than in partial cross-sections. It is seen that the action of $3d_{3/2}$ electrons leads to an extra maximum in the $3d_{5/2}$ cross-section.

It is interesting but perhaps incidental that the position of the main maxima in free Xe and Xe@$C_{60}$, just as in free Cs and Cs@$C_{60}$ almost coincide, but the entire endohedral atoms structure is much more complex.

Fig. 3a, b, c demonstrates the noticeable modifications in the dipole angular anisotropy parameter $\beta(\omega)$. However, the general behavior of $\beta(\omega)$, which in essence is a ratio (see (4) and (24)) of dipole matrix elements, is for endohedrals and of isolated atoms similar. Within the whole range of photon energy under consideration the curves $\beta(\omega)$ for the endohedral atoms Xe@$C_{60}$, Cs@$C_{60}$, and Ba@$C_{60}$ oscillate relative to the atomic curve. The fullerene shell especially vividly changes the frequency dependence of the parameter near the photoeffect threshold, where additional structure appears that is most impressive for Xe@$C_{60}$ and ionization of the 3/2 electrons in Ba. The additional

maximum of the curve $\beta_{5/2}(\omega)$ for photon energy $\omega \approx 51.5\,Ry$ exceeds by almost one and a half the maximum of the curve $\beta_{3/2}(\omega)$. Note, that reflection by $C_{60}$ generates in $\beta_{3/2}(\omega)$ a maximum that simulates in the maximum caused by intra-doublet correlations in $\beta_{5/2}(\omega)$ Some additional structure is created due to confinement of Xe inside $C_{60}$.

Figures 4a, b, c and 5a, b, c present the non-dipole parameters $\gamma^C(\omega)$ and $\delta^C(\omega)$, respectively, determined using equations (5), (6), (25), (26) and (3). The dashed lines on these figures give the data for isolated atoms, while the solid lines correspond to the endohedral atoms Xe@$C_{60}$, Cs@$C_{60}$, and Ba@$C_{60}$. As before, the upper curves correspond to ionization of the 5/2, while the lower to the 3/2 electrons. In Fig. 4 and 5 we also observe the oscillations of the frequency dependence parameters, which are especially vividly near the process threshold where the behavior of the background parameters for the 3/2 and 5/2 electrons are different as well. Interplay of intra-doublet resonances with confinement ones leads to strong interference phenomena that we can see in these figures. As a result of this interplay, the shapes of $\gamma^C_{5/2}(\omega)$ ( $\delta^C_{5/2}(\omega)$ ), and $\gamma^C_{3/2}(\omega)$ ( $\delta^C_{3.2}(\omega)$ ) for Xe@$C_{60}$ and Cs@$C_{60}$ (Fig.4a and 4b, Fig.5a and 5b), respectively, became closer than in isolated atoms where $\gamma^C_{5/2}(\omega)$ ( $\delta^C_{5/2}(\omega)$ ) has and additional maximum, while $\gamma^C_{3/2}(\omega)$ ( $\delta^C_{3.2}(\omega)$ ) has not.

Entirely, we see that the presence of the $C_{60}$ shell leads to prominent extra additional resonance structure in all the characteristics of the photoionization of Xe@$C_{60}$, Cs@$C_{60}$, and Ba@$C_{60}$ $3d_{5/2,3/2}$ electrons. It justifies studies of the photoionization of considered here endohedrals in the future.

It was discovered recently [19] that in accord with predictions in [3] for Cs and Ba the intra-doublet resonance in Cs is even stronger than in Xe. This serves as a qualitative confirmation of the presented above results of calculations, according to which the resonance structure in the $3d_{5/2}$ and $3d_{3/2}$ cross-sections of Cs@$C_{60}$ and Ba@$C_{60}$ is even stronger than in Xe@$C_{60}$.

## Acknowledgement

MYaA is grateful for financial support to the Israeli Science Foundation, Grant 174/03 and the Hebrew University Intramural Funds. ASB expresses his gratitude to the Hebrew University for hospitality and for financial support by Uzbekistan National Foundation, Grant Ф-2-1-12.

**References**

1. M. Ya. Amusia, A. S. Baltenkov, L. V. Chernysheva, Phys. Rev. A75, 043201 (2007).
2. A. Kivimäki, U. Hergenham, B. Kempgens, R. Hentges, M. N.Piancastelli, K. Maier, A. Ruedel, J. J. Tulkki, and B. M. Bradshaw, Phys. Rev. A 63, 2001, p. 012716.
3. M. Ya. Amusia, A. S. Baltenkov, L. V. Chernysheva, Z. Felfli, and A. Z. Msezane, J. Phys. B: At. Mol. Opt. Phys, **38**, L169-73, 2005.
4. M. J. Pushka and R. M. Niemenen. Phys. Rev. B **47**, 1181 (1993).
5. G. Wendin and B. Wastberg. Phys. Rev. B **48**, 14764 (1993).
6. L. S. Wang, J. M. Alford, Y. Chai, M. Diener, and R. E. Smalley. Z. Phys. D. **26**, S297 (1993).
7. P. Decleva, G. De Alti, M. Stener. J. Phys. B **32**, 4523 (1999).
8. J.-P. Connerade, V. K. Dolmatov, and S. T. Manson. J. Phys. B **33**, 2279 (2000).
9. J. P. Connerade, V. K. Dolmatov, and S. T. Manson J. Phys. B **33**, L275 (2000).
10. H. Shinohara. Rep. Prog. Phys. **63**, 843 (2000).
11. M. Stener, G. Fronzoni, D. Toffoli, P. Colavita, S. Furlan, and P. Decleva. J. Phys. B **35**, 1421 (2002).
12. J. Kou, T. Mori, M. Ono, Y. Haruyama, Y. Kubozono, and K. Mitsuke. Chem. Phys. Lett. **374**, 1 (2003).
13. R. Phaneuf, private communication (2007).
14. J.-P. Connerade, A. V. Solov'yov. J. Phys. B **3**8, 807 (2005).
15. M. Ya. Amusia, A. S. Baltenkov. Phys. Rev. A **73**, 062723 (2006).
16. M. Ya. Amusia, A. S. Baltenkov. Phys. Lett. A **360**, 294 (2006).
17. M. Ya. Amusia, L. V. Chernysheva, S. T. Manson, A. Z. Msezane and V. Radojevich, Phys. Rev. Lett. **88**, 093002 (2002).
18. V. Radojevich, D. Davidovich, and M. Ya. Amusia, Phys. Rev. A **67**, 22719-1-6 (2003).
19. H. Farrokhpour, M. Alagia, L. Avaldi, M. Ya. Amusia, L. V. Chernysheva, M. Coreno, M. de Simone, R. Richter, and S. Stranges), J. Phys. B: At. Mol. Opt. Phys. 39 765-771, 2006.
20. M. Ya. Amusia, Radiation Physics and Chemistry **70**, 237 (2004).
21. M. Ya. Amusia, P. U. Arifov, A. S. Baltenkov, A. A. Grinberg, and S. G. Shapiro, Phys. Lett. **47A**, 66 (1974).
22. M. Ya. Amusia, A. S. Baltenkov, L.V. Chernysheva, Z. Felfli, and A. Z. Msezane, Phys. Rev. A **63**, 052506 (2001).
23. J. W. Cooper, Phys. Rev. A **42**, 6942 (1990); Phys. Rev. A **45**, 3362 (1992); Phys. Rev. A **47**, 1841 (1993).
24. A. Bechler and R. H. Pratt, Phys. Rev. A **42**, 6400 (1990).
25. M. Ya. Amusia, *Atomic Photoeffect* (Plenum Press, New York – London, 1990).
26. M. Ya. Amusia and L.V. Chernysheva, *Computation of Atomic Processes* ("Adam Hilger" Institute of Physics Publishing, Bristol – Philadelphia, 1997).
27. M. Ya. Amusia, A. S. Baltenkov and B. G. Krakov, Phys. Lett. A **243**, 99 (1998).
28. M. Ya. Amusia, A. S. Baltenkov, L. V. Chernysheva, Z. Felfli, S. T. Manson, and A. Z. Msezane, J. Phys. B: At. Mol. Opt. Phys., **37**, 937-944, 2004
29. M. Ya. Amusia, A. S. Baltenkov, V. K. Dolmatov, S. T. Manson, and A. Z. Msezane, Phys. Rev. A **70**, 023201 (2004).

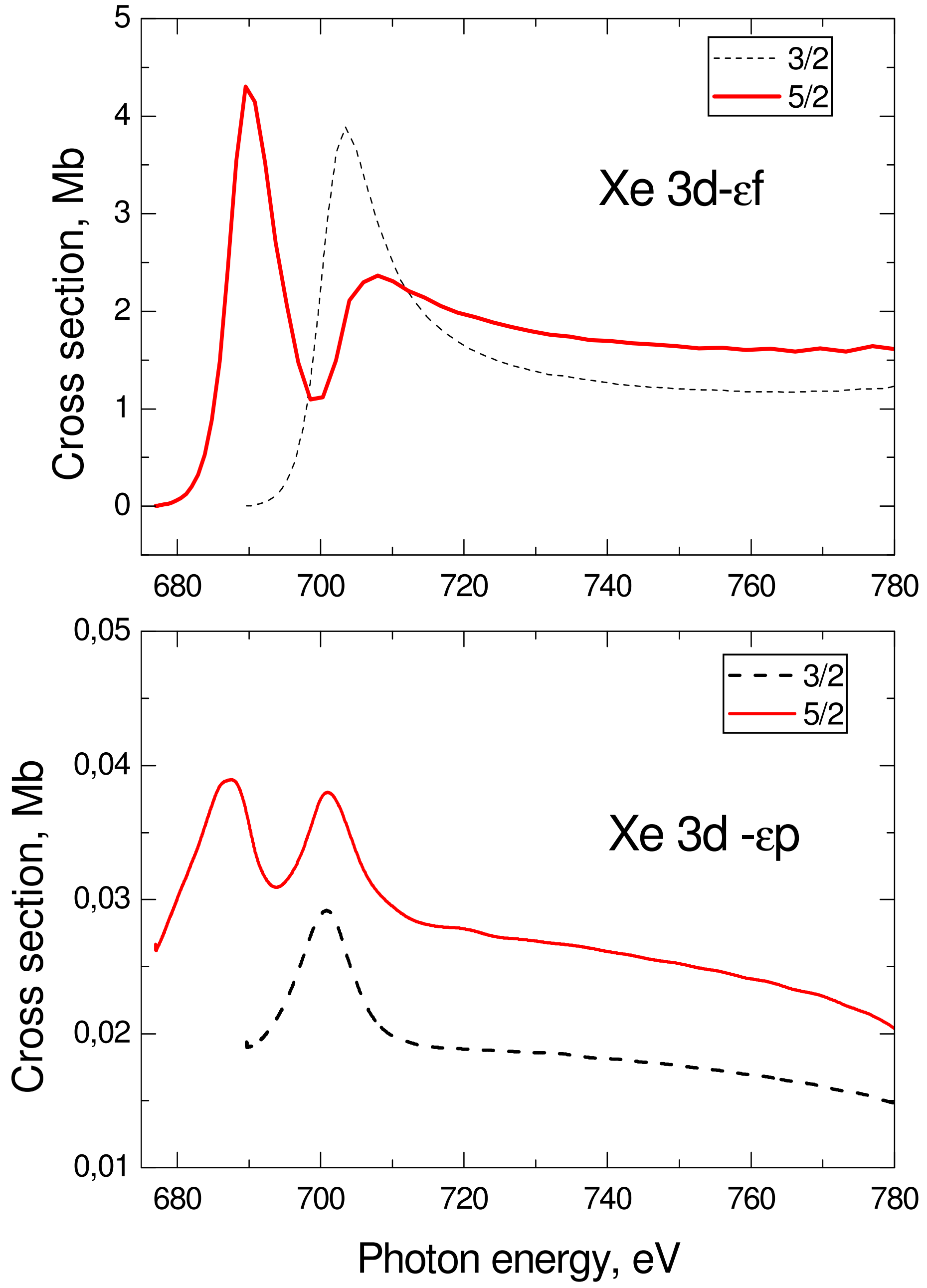

Fig.1a. Photoionization cross-sections of Xe 3*d* electrons, $3d-\varepsilon f$ and $3d-\varepsilon p$ transitions.

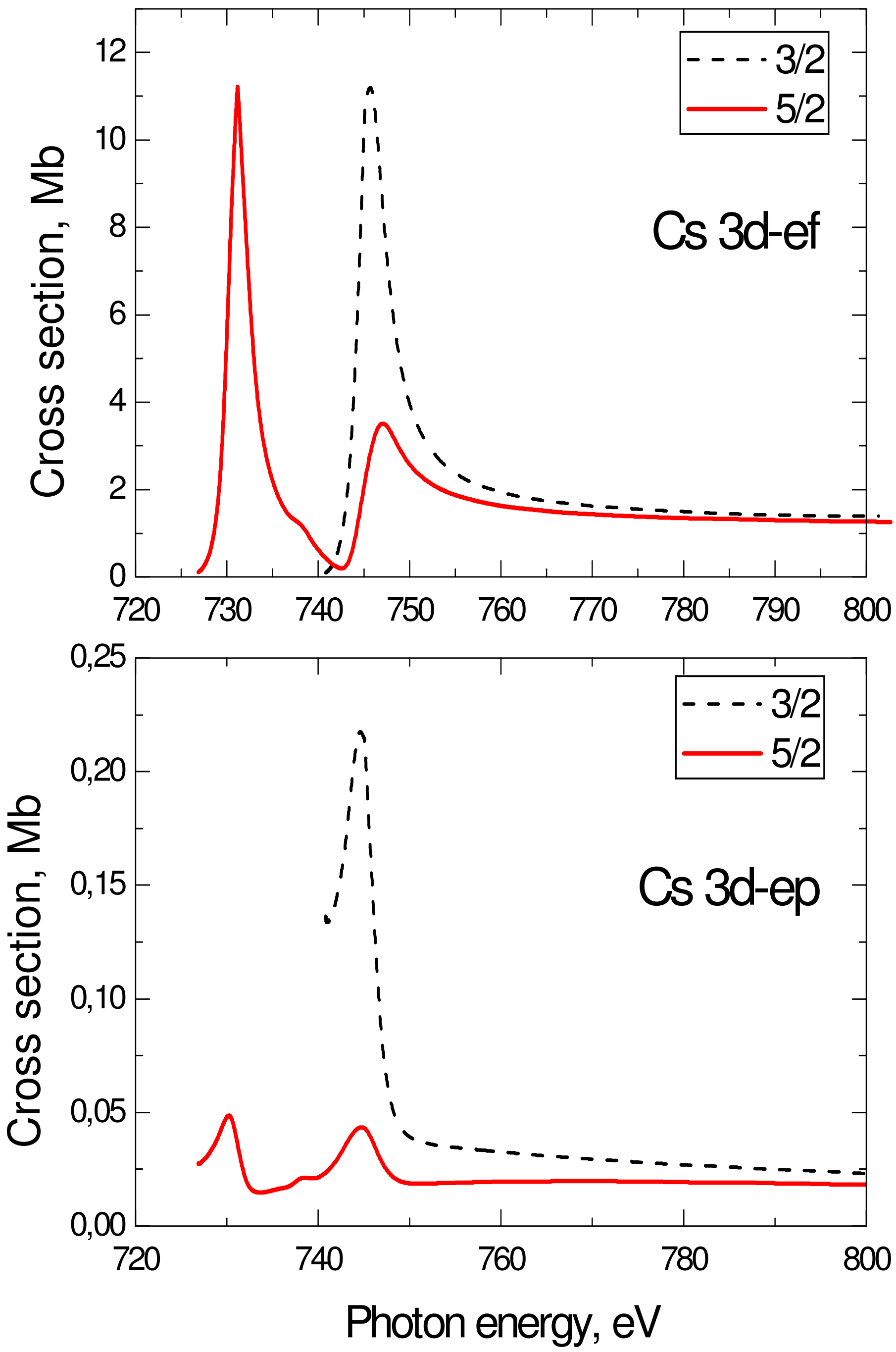

Fig.1b. Photoionization cross-sections of Cs 3*d* electrons, $3d-\varepsilon f$ and $3d-\varepsilon p$ transitions.

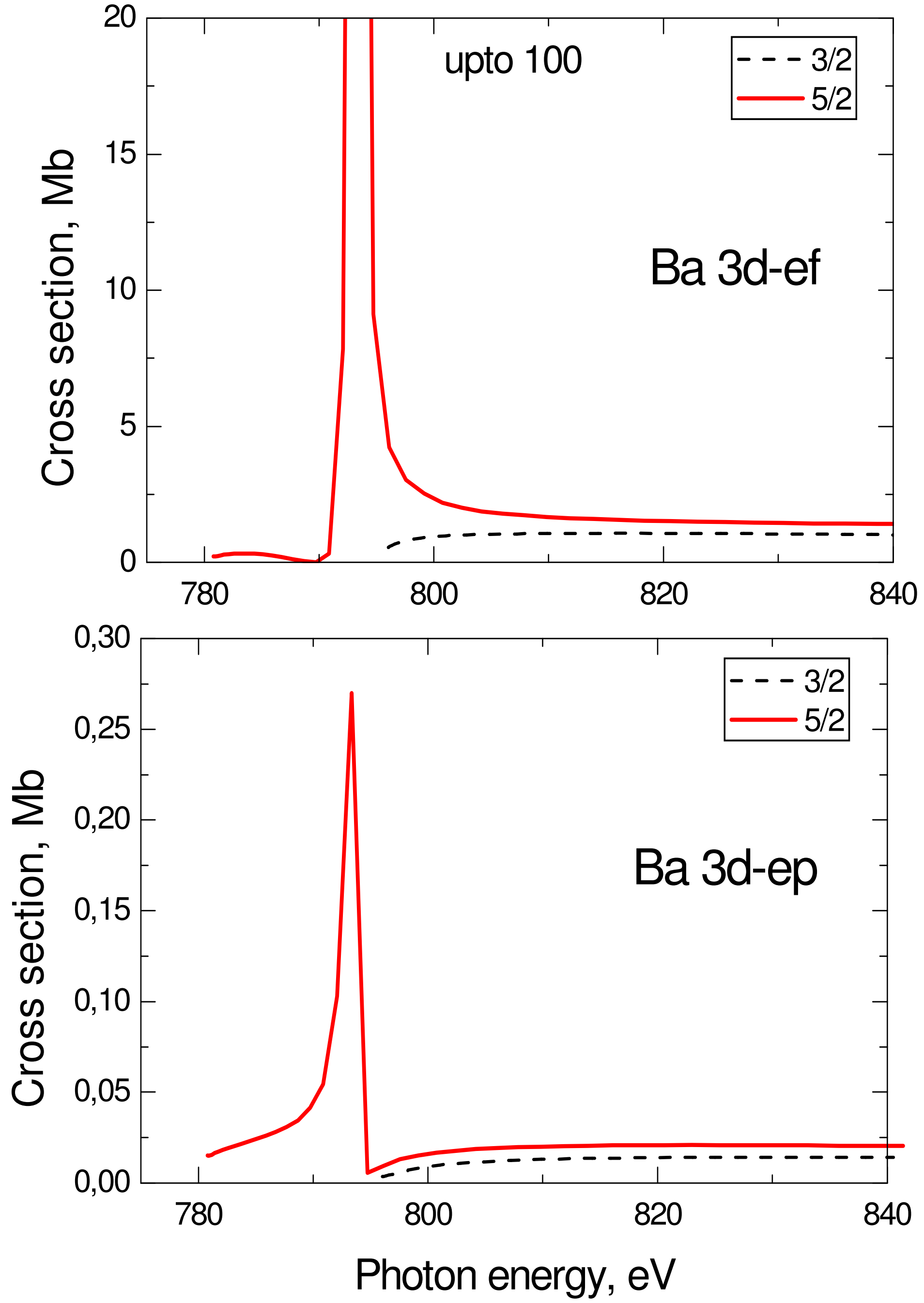

Fig.1c. Photoionization cross-sections of Ba 3*d* electrons,
$3d-\varepsilon f$ and $3d-\varepsilon p$ transitions.

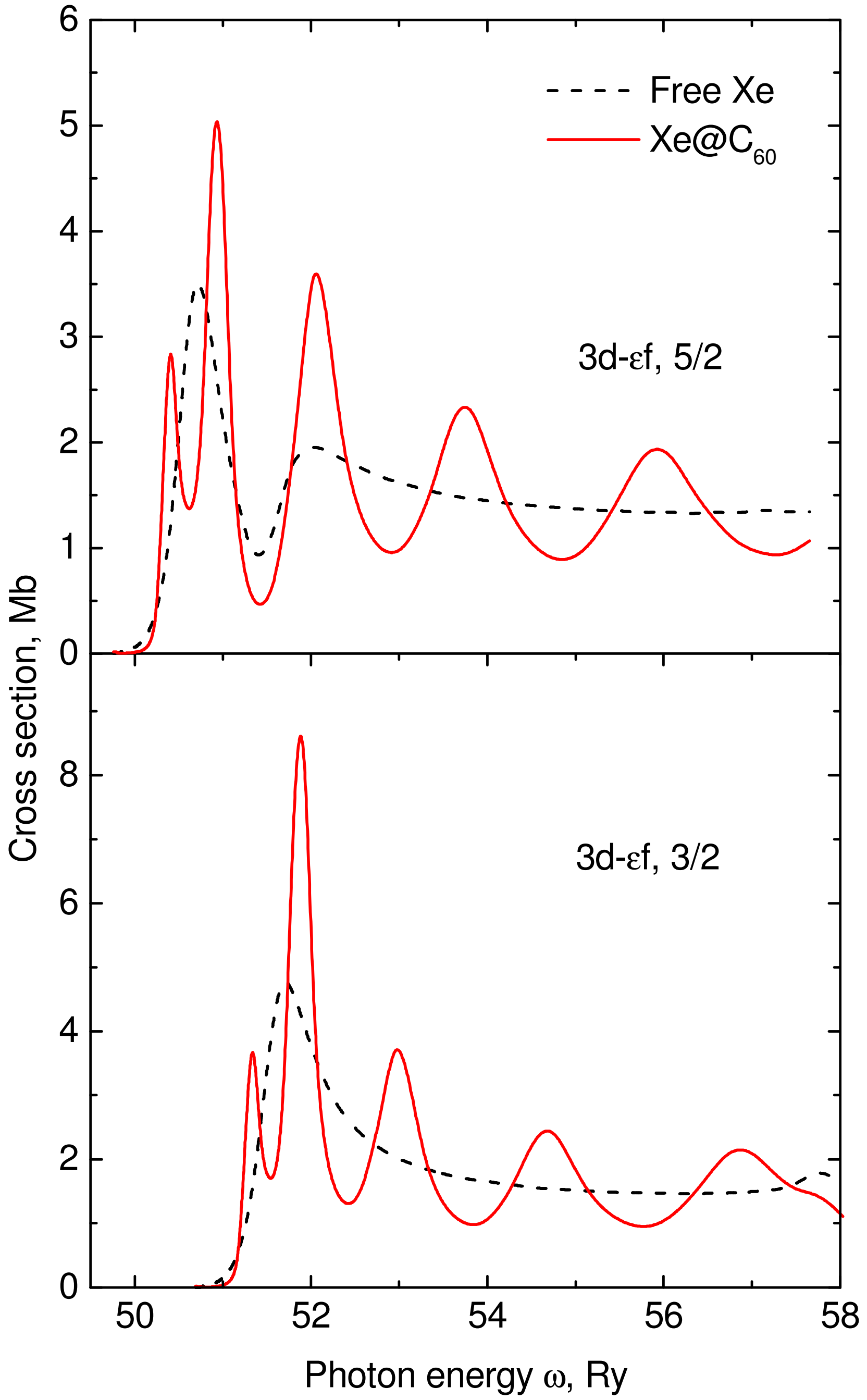

Fig. 2a. Photoionization cross-sections of $3d_{3/2}$ and $3d_{5/2}$ electrons in Xe and Xe@$C_{60}$.

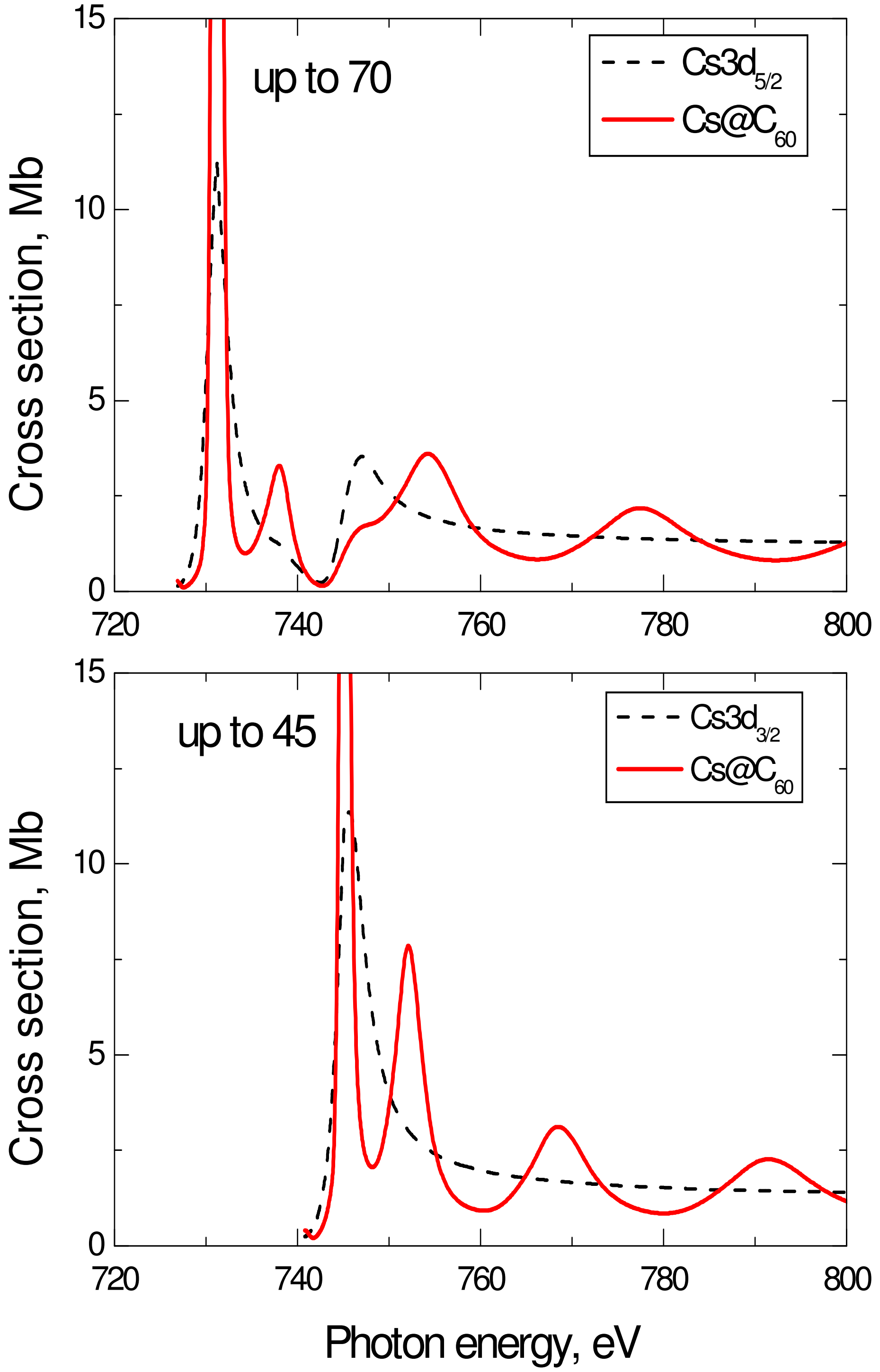

Fig. 2b. Photoionization cross-sections of $3d_{3/2}$ and $3d_{5/2}$ electrons in Cs and $Cs@C_{60}$.

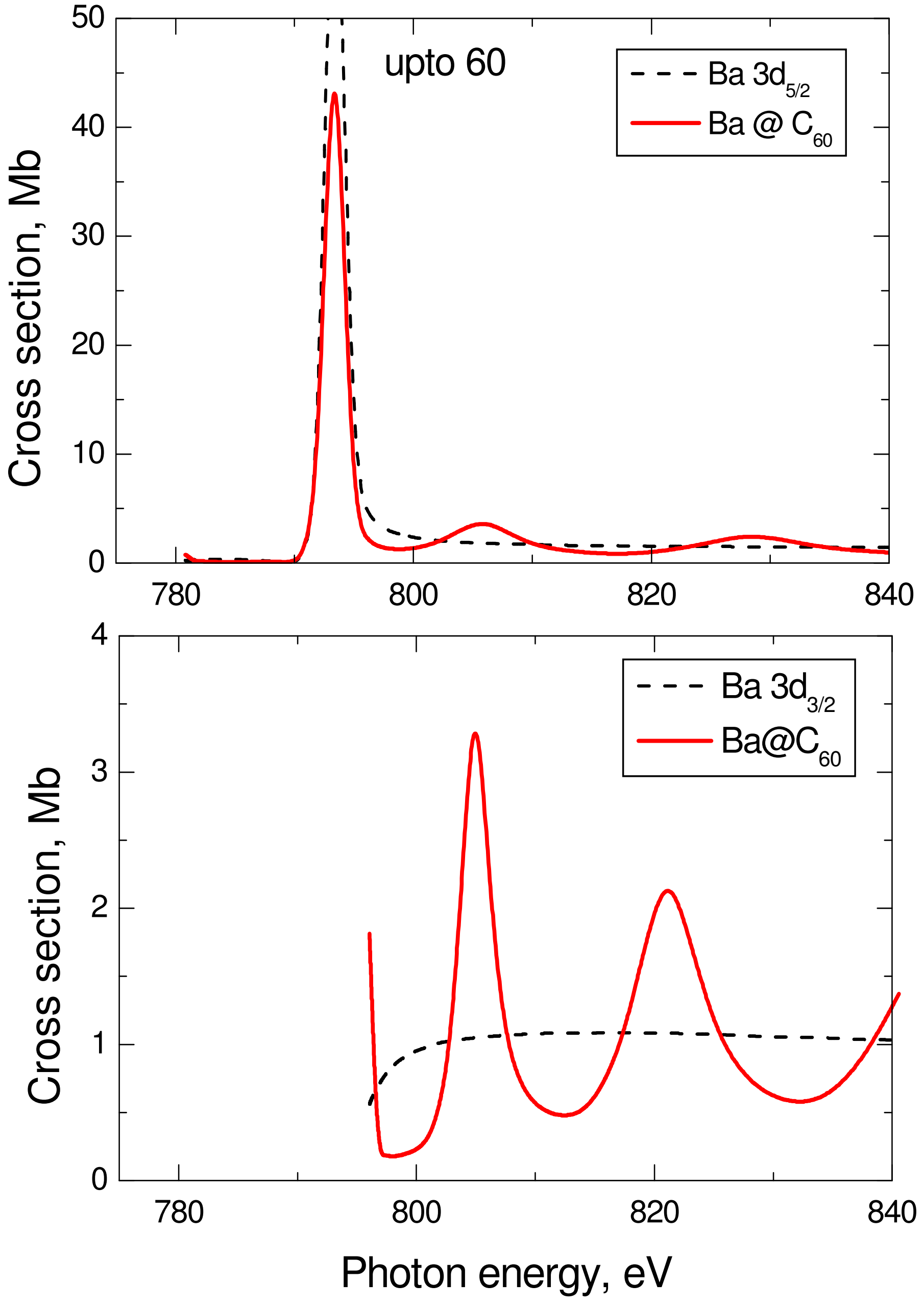

Fig. 2c. Photoionization cross-sections of $3d_{3/2}$ and $3d_{5/2}$ electrons in Ba and Ba@$C_{60}$.

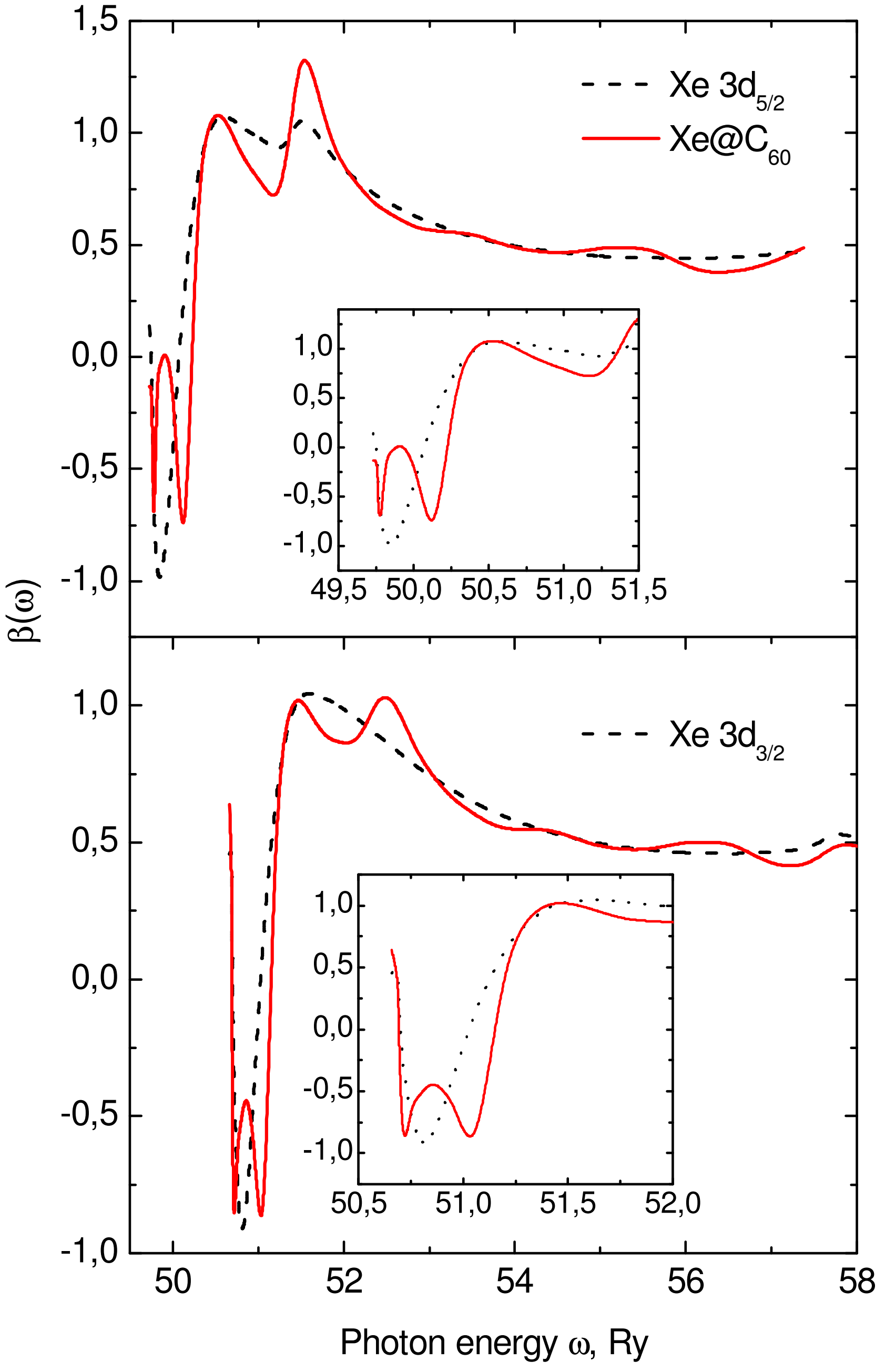

Fig. 3a. Dipole angular anisotropy parameter $\beta(\omega)$ for $3d_{3/2}$ and $3d_{5/2}$ electrons of Xe. The solid lines are the result for the endohedral Xe@$C_{60}$. The corresponding curves near the photoionization threshold are presented in the insets.

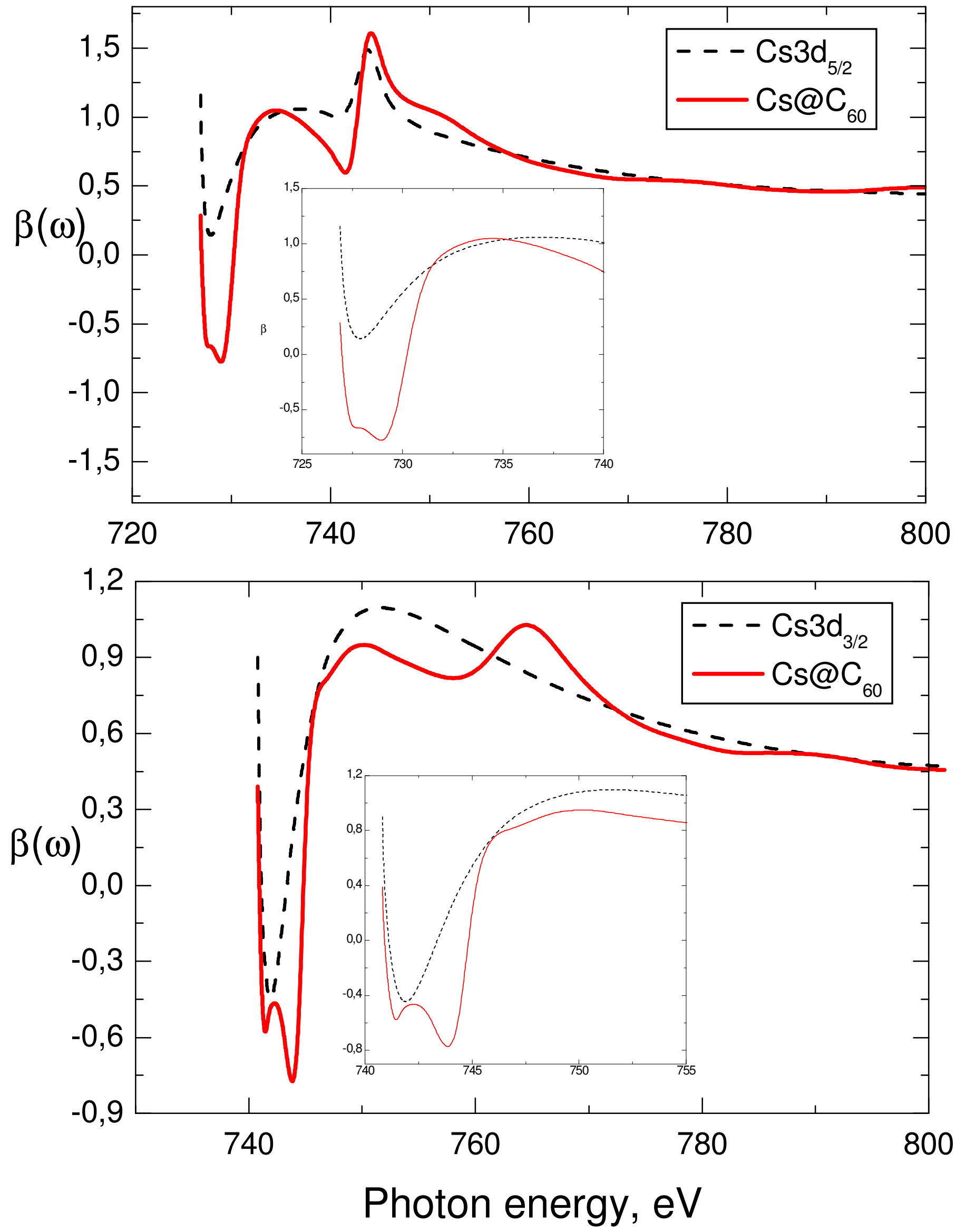

Fig. 3b. Dipole angular anisotropy parameter $\beta(\omega)$ for $3d_{3/2}$ and $3d_{5/2}$ electrons of Cs. The solid lines are the calculation result for the endohedral systems $Cs@C_{60}$. The corresponding curves near the photoionization threshold are presented in the insets.

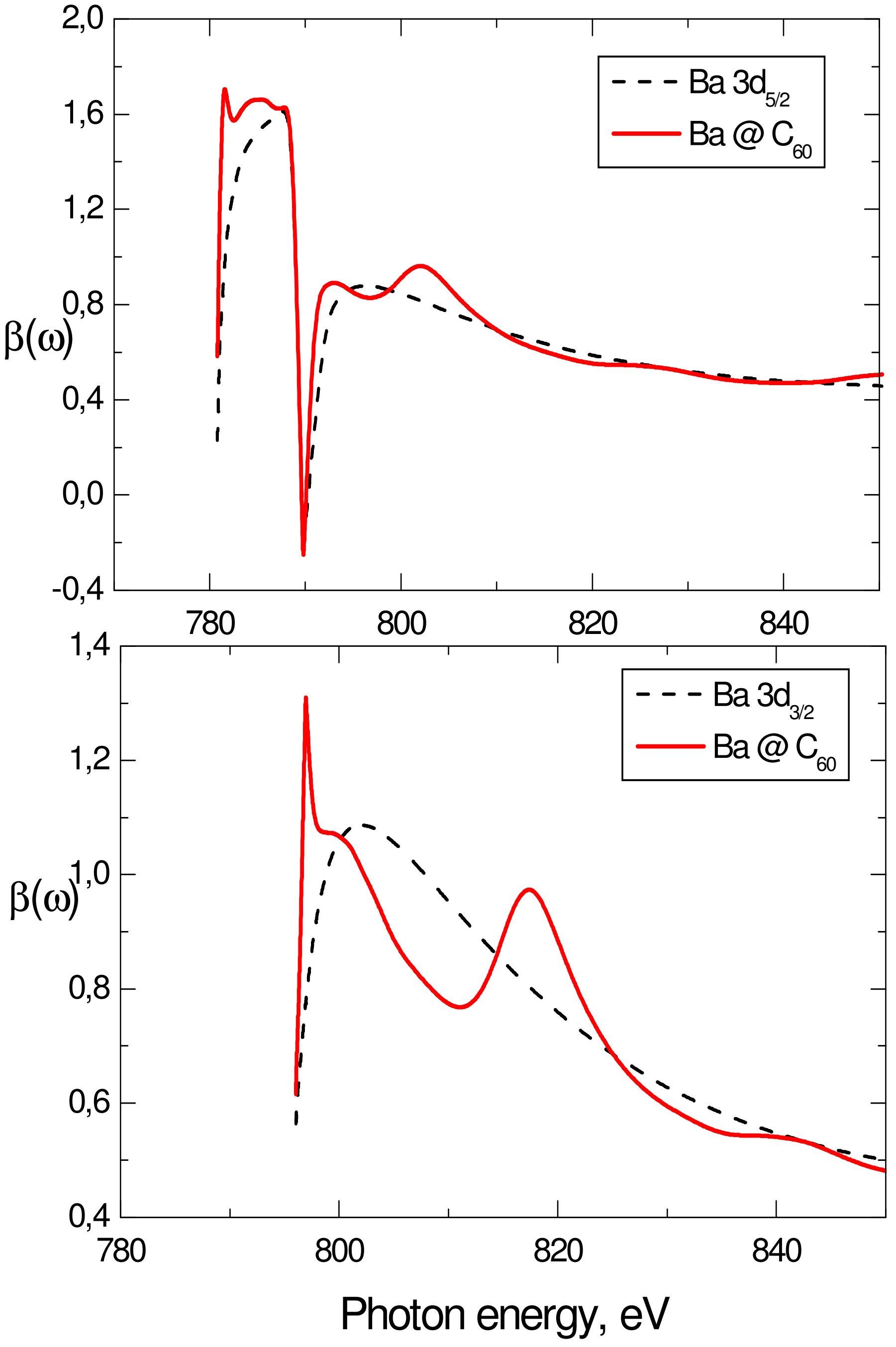

Fig. 3c. Dipole angular anisotropy parameter $\beta(\omega)$ for $3d_{3/2}$ and $3d_{5/2}$ electrons of Ba. The solid lines are the calculation result for the endohedral systems Ba@$C_{60}$.

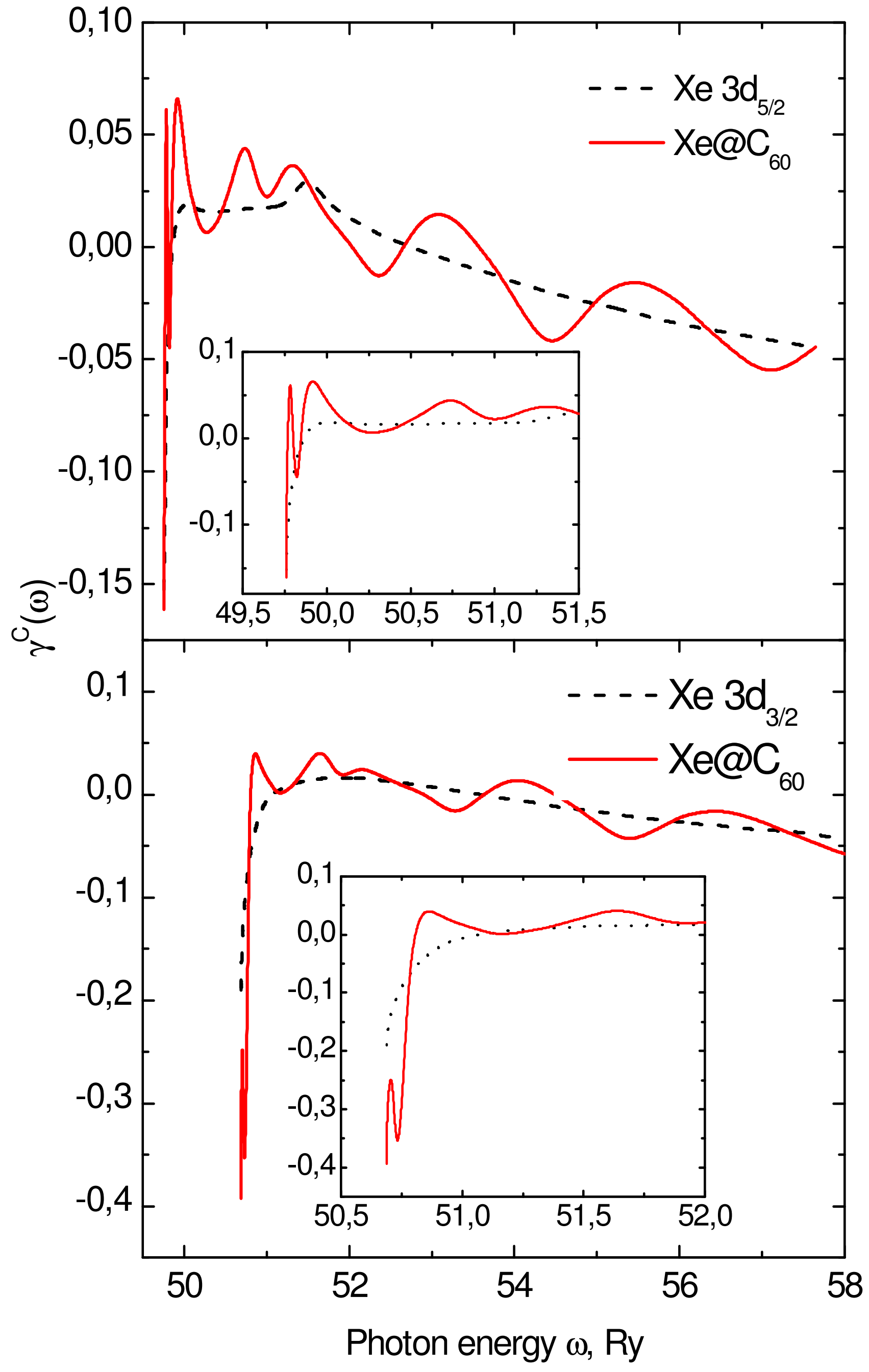

Fig. 4a. Non-dipole angular anisotropy parameter $\gamma^C(\omega)$ for $3d_{3/2}$ and $3d_{5/2}$ electrons of Xe. The solid lines are the result for the endohedral $Xe@C_{60}$. The corresponding curves near the photoionization threshold are presented in the insets.

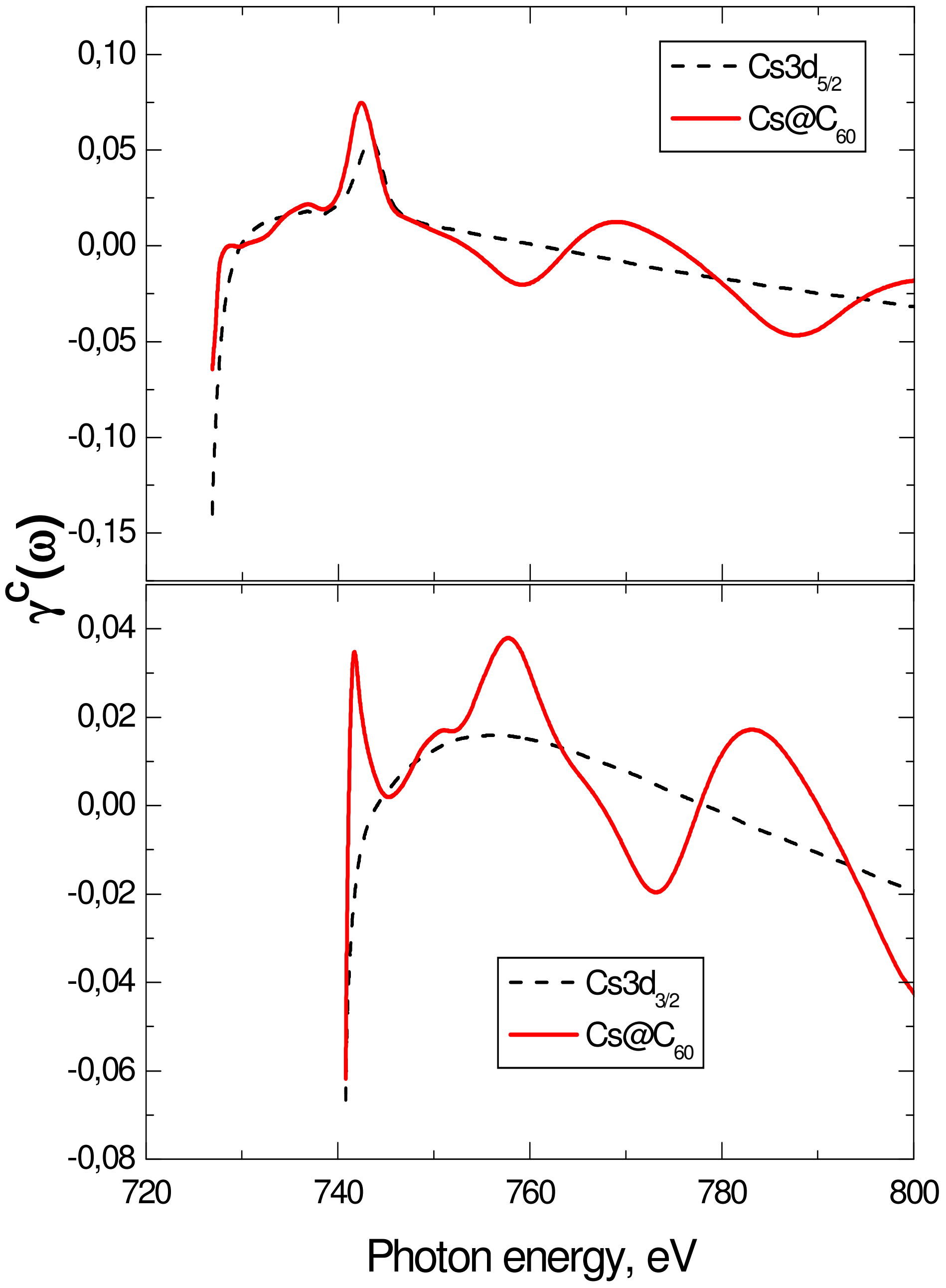

Fig. 4b. Non-dipole angular anisotropy parameter $\gamma^{C}(\omega)$ for $3d_{3/2}$ and $3d_{5/2}$ electrons of Cs. The solid lines are the result for the endohedral $Cs@C_{60}$.

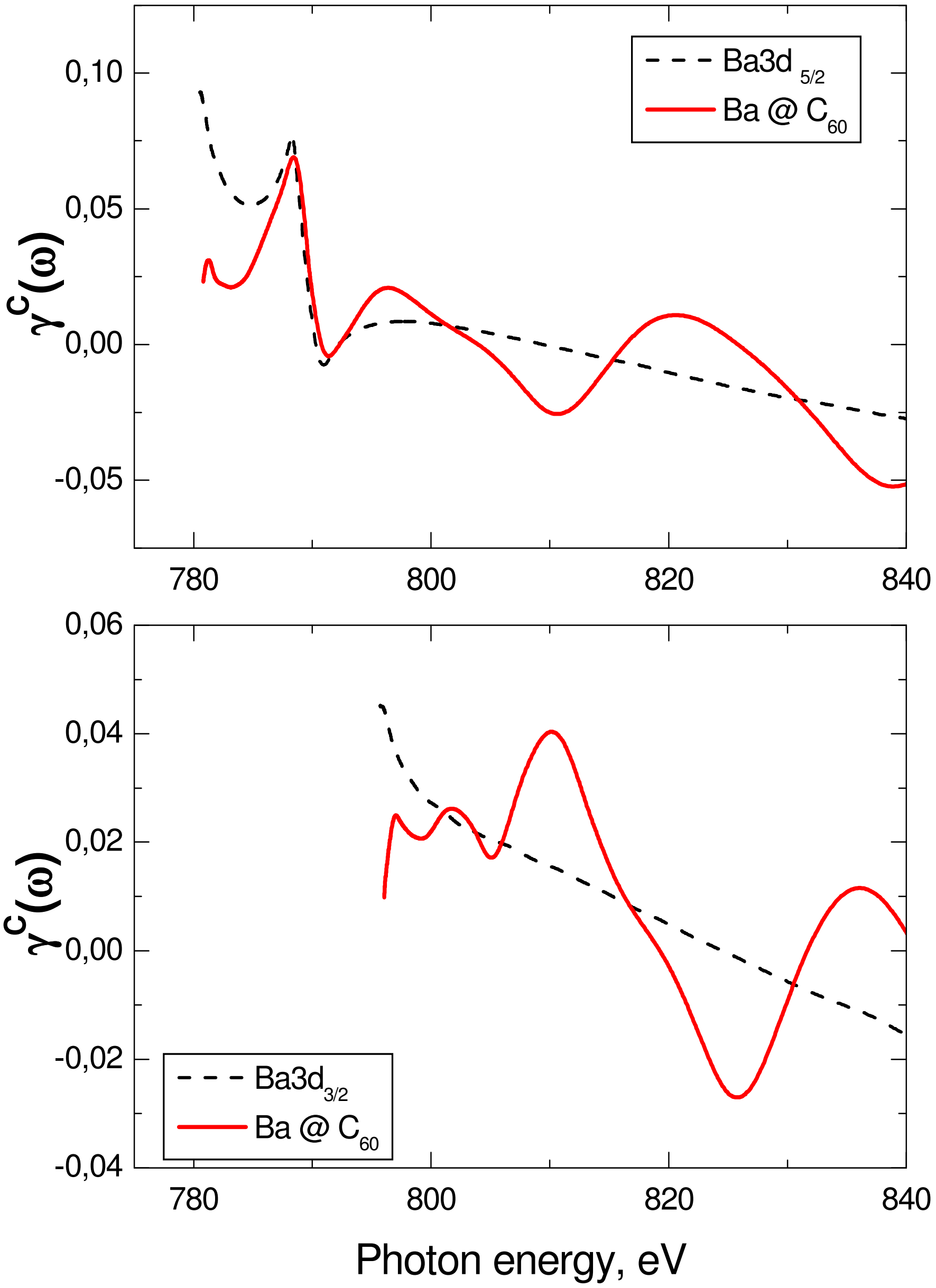

Fig. 4c. Non-dipole angular anisotropy parameter $\gamma^{C}(\omega)$ for $3d_{3/2}$ and $3d_{5/2}$ electrons of Ba. The solid lines are the result for the endohedral $Ba@C_{60}$.

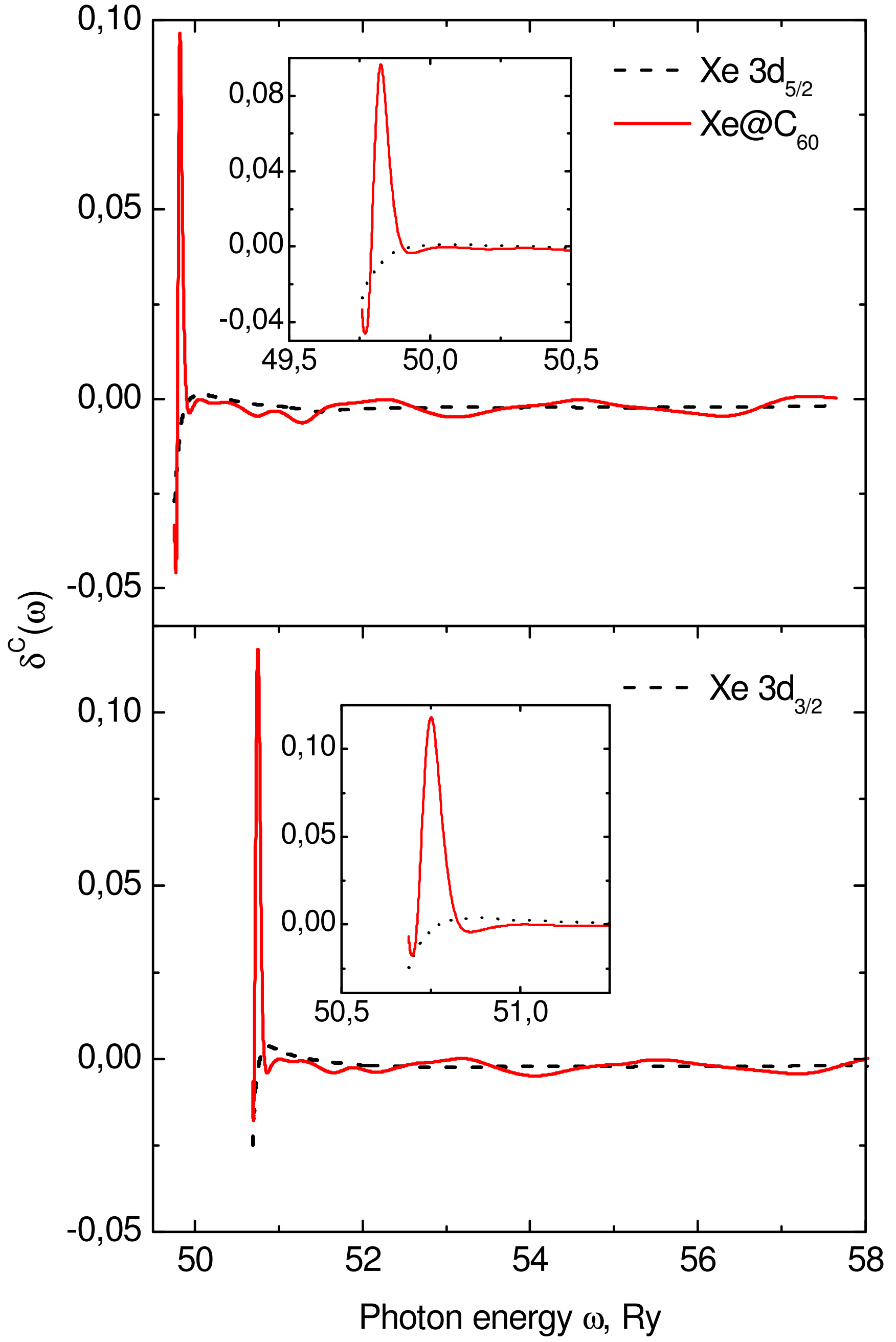

Fig. 5a. Non-dipole angular anisotropy parameter $\delta^{C}(\omega)$ for $3d_{3/2}$ and $3d_{5/2}$ electrons of Xe. The solid lines are the result for the endohedral $Xe@C_{60}$. The corresponding curves near the photoionization threshold are presented in the insets.

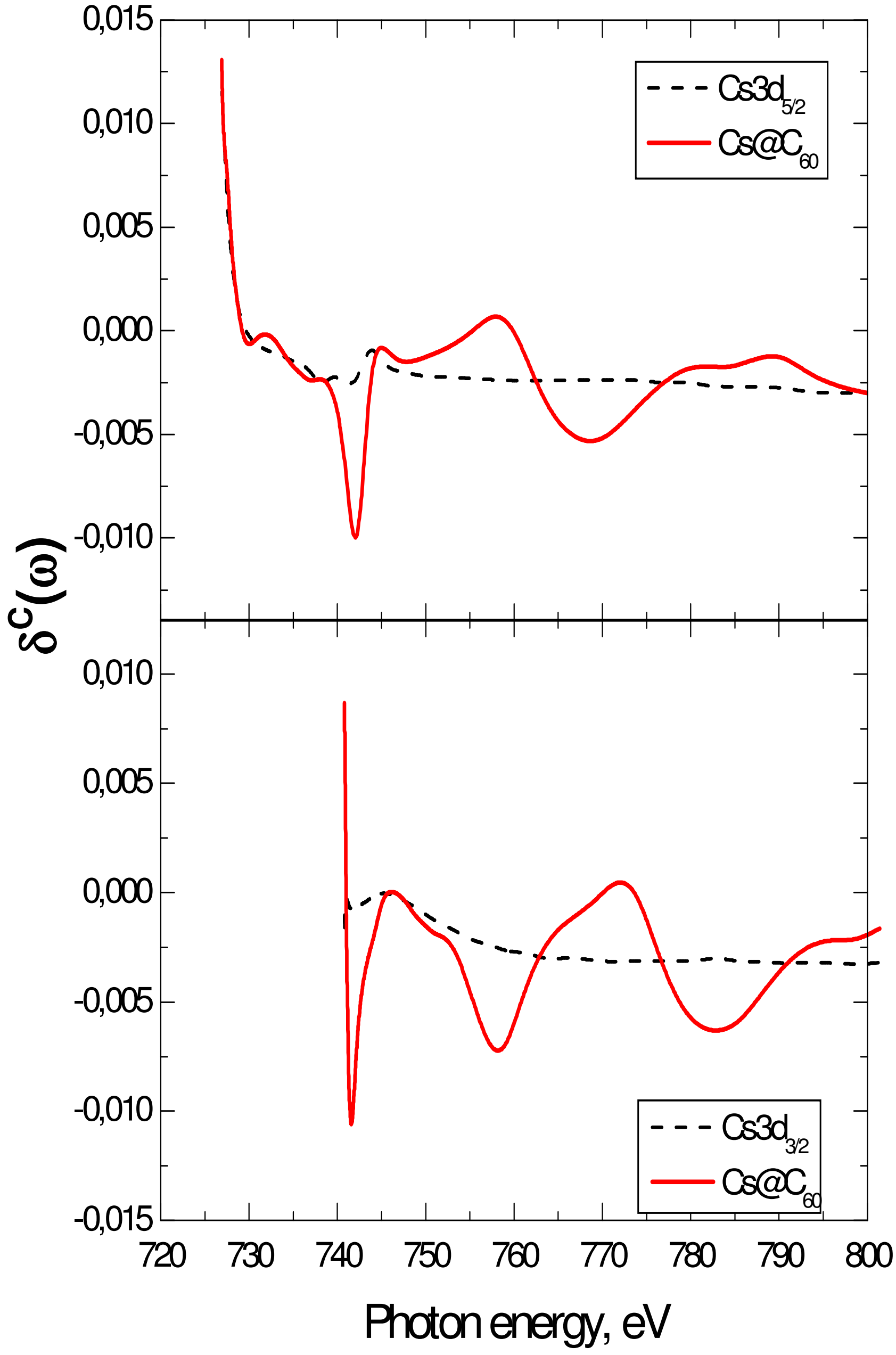

Fig. 5b. Non-dipole angular anisotropy parameter $\delta^{C}(\omega)$ for $3d_{3/2}$ and $3d_{5/2}$ electrons of Cs. The solid lines are the result for the endohedral $Cs@C_{60}$.

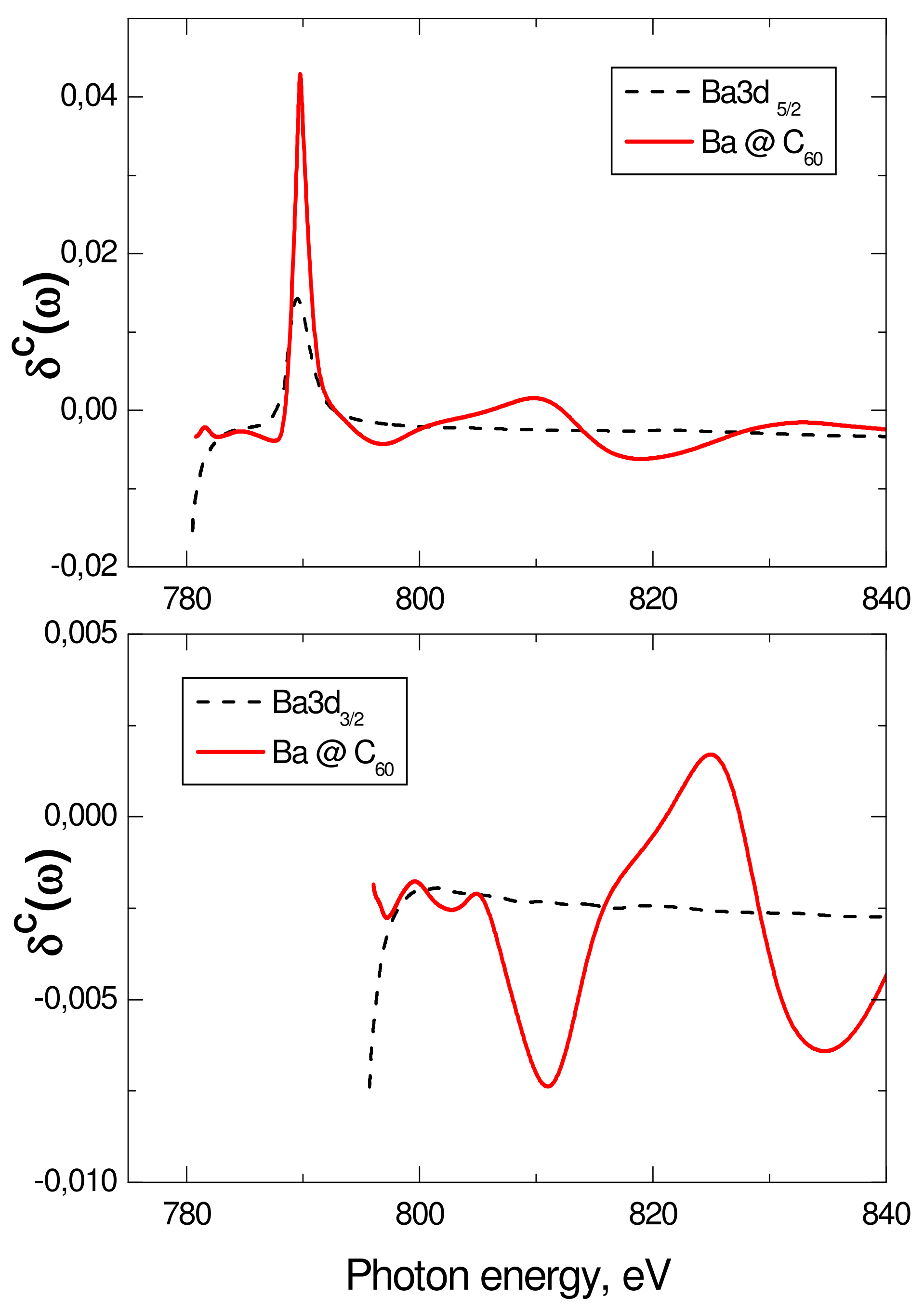

Fig. 5c. Non-dipole angular anisotropy parameter $\delta^C(\omega)$ for $3d_{3/2}$ and $3d_{5/2}$ electrons of Ba. The solid lines are the result for the endohedral $Ba@C_{60}$.